\documentclass[aps,prd,preprint,groupedaddress,preprintnumbers]{revtex4-1}

\usepackage{graphicx}
\newcommand{\nn}{\nonumber}
\newcommand{\e}{\hbox{e}}
\newcommand{\Veff}{V_{\mathrm{eff}}}
\newcommand{\vev}[1]{\left\langle #1 \right\rangle}
\newcommand{\mph}{m^2_{\mathrm{ph}}}
\newcommand{\xph}{\frac{\lambda_{\mathrm{ph}}}{(4 \pi)^2}}
\newcommand{\lb}{\left\lbrace}
\newcommand{\rb}{\right\rbrace}
\begin{document}

\preprint{KOBE-TH-13-04}

\title{Analytic form of the effective potential in the large $N$ limit
of a real scalar theory in four dimensions}


\author{H.~Sonoda}
\email[]{hsonoda@kobe-u.ac.jp}
\affiliation{Physics Department, Kobe University, Kobe 657-8501 Japan}


\date{13 March 2013}

\begin{abstract}
    We give the large $N$ limit of the effective potential for the
    $O(N)$ linear sigma model in four dimensions in terms of the
    Lambert $W$ function.  The effective potential is fully consistent
    with the renormalization group, and it admits an asymptotic
    expansion in powers of a small positive coupling parameter.
    Careful consideration of the UV cutoff present in the model
    validates the physics of the large $N$ limit.
\end{abstract}


\maketitle



\section{Introduction}

The physics of the $O(N)$ linear sigma model in four dimensions is
well known.  The model has two phases: the symmetric phase where the
$N$ scalar fields obtain the same mass, and the broken phase where the
spontaneous breaking of $O(N)$ to $O(N-1)$ gives rise to $N-1$
massless Nambu-Goldstone bosons.  In the large $N$ limit, the model
becomes weakly coupled and can be studied in
details:\cite{Coleman:1974jh, Coleman:1985, Moshe:2003xn,
  Coleman:1985} the physical meaning of the triviality can be
elucidated, and the physical equivalence between the linear and
non-linear versions of the model, for a large enough UV cutoff, can be
demonstrated explicitly.  For three dimensions, an analytic form of
the large $N$ limit of the effective potential has been known for a
long time (see, for example, \cite{Coleman:1974jh, David:1985zz}), and
the existence of the non-trivial Wilson-Fisher fixed point has also
been shown explicitly.  In four dimensions, however, such an analytic
form of the effective potential has not been given to the author's
knowledge.

It is the purpose of this short paper to give an exact analytic form of
the effective potential in the large $N$ limit of the $O(N)$ linear
sigma model in four dimensions.  The obtained effective potential is not
only fully consistent with the renormalization group (RG), but it also
illuminates the nature of asymptotic expansions in powers of a small
coupling.  Since we assume the smallness of the physical mass compared
with the UV cutoff, our effective potential is valid only for the VEV of
a scalar field much smaller than the cutoff.

At the end of the paper, we have added a section to defend the physics
of the large $N$ limit.  We resolve the two problems raised in
\cite{Coleman:1974jh} by taking into account the presence of a UV
cutoff.

\section{The large $N$ limit of the $O(N)$ linear sigma model}

We consider the $O(N)$ linear sigma model in four dimensional
Euclidean space.  The effective potential for the large $N$ limit has
been determined implicitly in \cite{Coleman:1974jh}.  We extend the
result to give an explicit analytic form of the effective potential
using the Lambert $W$ function.\cite{wiki:LambertW}

The action is given by
\begin{equation}
S[\phi] = \int d^4 x\, \left[ \frac{1}{2} \left(\partial_\mu \phi^I
    \partial_\mu \phi^I \right) + \frac{m_0^2}{2} \left(\phi^I \phi^I\right) +
    \frac{\lambda_0}{8 N} \left(\phi^I \phi^I\right)^2 \right]
\end{equation}
where the repeated $I$ is summed over $I= 1, \cdots, N$.  We introduce
a constant source $j$ and define a generating function $w(j)$ by
\begin{equation}
\e^{- N w (j) \int d^4 x} = \int \left[d\phi^I\right] \,
\exp \left[- S [\phi] + \sqrt{N} j \int d^4 x\, \phi^N (x) \right]
\end{equation}
so that
\begin{equation}
- w' (j) = v \equiv \frac{1}{\sqrt{N}} \vev{\phi^N}\label{wprime}
\end{equation}
We then define the effective potential by the Legendre transform
\begin{equation}
\Veff (v) \equiv w(j) + v j
\end{equation}
To compute $\Veff$, we invert (\ref{wprime}) to give $j$ in terms of
$v$.  We then integrate
\begin{equation}
\Veff' (v) = j
\end{equation}
to obtain $\Veff$.

The method of the large $N$ limit is well known.\cite{Coleman:1974jh,
  Coleman:1985, Moshe:2003xn} We first introduce an auxiliary field
$\alpha (x)$ to rewrite the action with source as
\begin{eqnarray}
S [\phi, \alpha] &=& \int d^4 x \Bigg[
\frac{1}{2} \left(\partial_\mu \phi^I \partial_\mu \phi^I\right) +
\frac{m_0^2}{2} \left(\phi^I \phi^I\right) + \frac{\lambda_0}{8 N}
\left( \phi^I \phi^I \right)^2 \nn\\
&&\quad + \frac{N}{2 \lambda_0} \left( \alpha + i \frac{\lambda_0}{2
          N} \left( \phi^I \phi ^I \right) \right)^2 - \sqrt{N} j
    \phi^N \Bigg]\nn\\
&=& \int d^4 x \left[ \frac{1}{2} \left(\partial_\mu
        \phi^I \partial_\mu \phi^I\right) + 
\frac{1}{2} \left( m_0^2 + i \alpha\right) \left(\phi^I \phi^I\right) 
+ \frac{N}{2 \lambda_0} \alpha^2 - \sqrt{N} j
    \phi^N \right]
\end{eqnarray}
In the large $N$ limit, let $\Delta m_0^2$ be the saddle point value
of $i \alpha$, which is determined by
\begin{equation}
\frac{1}{2} \left( v^2 + \int \frac{d^4 p}{(2 \pi)^4} \frac{1}{p^2 +
      m_0^2 + \Delta m_0^2} 
\right) = \frac{\Delta m_0^2}{\lambda_0}\label{first}
\end{equation}
$\vev{\phi^N} = \sqrt{N} v$ gives
\begin{equation}
v\left( m_0^2 + \Delta m_0^2 \right)  = j = \Veff' (v) \label{second}
\end{equation}
Eqs.~(\ref{first}) and (\ref{second}) are Eqs.~(2.7) and (2.8) of
Ref.~\cite{Coleman:1974jh}.  In the following we first solve
(\ref{first}) to determine $m_0^2 + \Delta m_0^2$ in terms of $v$.  It
is then straightforward to obtain $\Veff (v)$ by integrating
(\ref{second}).

To begin with, we introduce
\begin{equation}
\mph \equiv m_0^2 + \Delta m_0^2\label{mph}
\end{equation}
which is the physical squared mass of the field $\phi^I$.  We assume
that it is non-negative, and that it is extremely small compared with
the square of the UV cutoff $\Lambda_0$:
\begin{equation}
0 \le \mph \ll \Lambda_0^2
\label{assumption}
\end{equation}
We then obtain
\begin{equation}
\int \frac{d^4 p}{(2 \pi)^4} \frac{1}{p^2 + \mph} = A_4 \Lambda_0^2 +
\frac{\mph}{(4 \pi)^2} \lb \ln \frac{\mph}{\Lambda_0^2} + B_4 \rb
+ \mathrm{O} \left( \left(\mph\right)^2/\Lambda_0^2\right)
\label{integral}
\end{equation}
where $A_4$ and $B_4$ are regularization dependent constants.  The
assumption (\ref{assumption}) allows us to ignore the corrections
inversely proportional to $\Lambda_0^2$.  Thus, (\ref{first}) gives
\begin{eqnarray}
m_0^2 &=& \mph - \Delta m_0^2\nn\\
&=& - \frac{\lambda_0}{2} A_4 \Lambda_0^2 +
\frac{\lambda_0}{(4 \pi)^2} \left[
\mph \left( \frac{(4\pi)^2}{\lambda_0} - \frac{1}{2} \ln
    \frac{\mph}{\Lambda_0^2} - \frac{1}{2} B_4 \right) - (4 \pi)^2
\frac{v^2}{2} \right]\label{m0squared}
\end{eqnarray}

Let us now introduce a renormalized coupling $\lambda$ by
\begin{equation}
\frac{\lambda}{(4 \pi)^2} \equiv \frac{1}{\ln \frac{\Lambda_0}{\mu} +
  \frac{(4\pi)^2}{\lambda_0} - \frac{1}{2} B_4}
\end{equation}
where $\mu$ is an arbitrary finite renormalization scale.  It
satisfies the RG equation
\begin{equation}
- \mu \frac{\partial}{\partial \mu} \frac{\lambda}{(4 \pi)^2} = -
\left( \frac{\lambda}{(4 \pi)^2} \right)^2
\end{equation}
We then introduce an RG invariant mass scale by
\begin{equation}
\Lambda \equiv \e^{\frac{(4\pi)^2}{\lambda}} \mu\label{Lambda}
\end{equation}
If $\frac{\lambda_0}{(4 \pi)^2}$ is of order $1$, $\Lambda$ is of the
same order as $\Lambda_0$.

Let us next introduce a squared mass parameter $m^2$ which is either
positive or negative:
\begin{equation}
m_0^2 = - \frac{\lambda_0}{2} A_4 \Lambda_0^2 +
\frac{\lambda_0}{\lambda} m^2\label{m2def}
\end{equation}
The first term on the right is the critical squared mass; the $O(N)$
symmetry is exact for $m^2 > 0$, and spontaneously broken to $O(N-1)$
for $m^2 < 0$.  The squared mass satisfies
\begin{equation}
- \mu \frac{\partial}{\partial \mu} m^2 = - \frac{\lambda}{(4 \pi)^2}
m^2
\end{equation}

Using (\ref{m2def}), we can rewrite (\ref{m0squared}) as
\begin{equation}
\mph \left( 1 - \frac{\lambda}{(4 \pi)^2} \frac{1}{2} \ln
    \frac{\mph}{\mu^2} \right) = m^2 + \frac{1}{2} \lambda v^2
\label{old}
\end{equation}
This is the same as (2.11) of Ref.~\cite{Coleman:1974jh}.  With
(\ref{Lambda}), this can be rewritten further as
\begin{equation}
\frac{\mph}{\Lambda^2} \ln \frac{\mph}{\Lambda^2} = \xi
\label{to-solve}
\end{equation}
where $\xi$ is short for the RG invariant
\begin{equation}
  \xi \equiv - 2
  \frac{\frac{(4\pi)^2}{\lambda} \left( m^2 + \frac{1}{2} \lambda v^2\right)}
    {\Lambda^2} = \frac{m^2 + \frac{1}{2} \lambda v^2}{\Lambda^2}
    \ln \frac{\mu^2}{\Lambda^{2}}
\label{xidef}
\end{equation}
Since $\mph \ll \Lambda^2 \simeq \Lambda_0^2$ by the assumption
(\ref{assumption}), we find
\begin{equation}
- \xi \ll 1
\end{equation}
This corresponds to
\begin{equation}
v^2 \ll \Lambda^2 
\label{smallv}
\end{equation}

Now, the solution of (\ref{to-solve}) is obtained by the lower branch
of the Lambert $W$ function (or the product logarithm):
\begin{equation}
\ln \frac{\mph}{\Lambda^2} = W_{-1} (\xi)
\end{equation}
where the negative valued $W_{-1} (\xi)$ is defined implicitly by
\begin{equation}
W_{-1} (\xi)\, \e^{W_{-1} (\xi)} = \xi
\end{equation}
for $\xi \in [-\frac{1}{\e}, 0)$.  (See Appendix for a little more
details on $W_{-1}$.)  

\begin{figure}[bh]
\includegraphics[width=12cm]{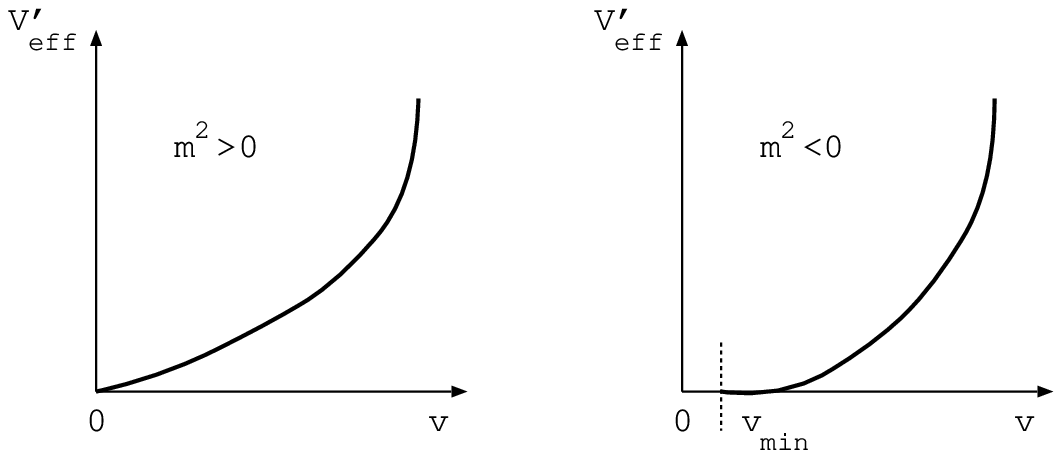}
\caption{For $m^2 > 0$, $\Veff$ is defined for $0 \le v \le
  \sqrt{\frac{1}{(4 \pi)^2} \Lambda^2 \e^{-1} - \frac{2
      m^2}{\lambda}}$.  For $m^2 < 0$, it is defined for
  $v_{\mathrm{min}}\equiv \sqrt{\frac{- 2 m^2}{\lambda}} < v <
  \sqrt{\frac{1}{(4 \pi)^2} \Lambda^2 \e^{-1} - \frac{2
      m^2}{\lambda}}$.  At $v=v_{\mathrm{min}}$, $\Veff' = 0$.}
\label{derivative}
\end{figure}

Using (\ref{second}) and (\ref{mph}), we thus obtain
\begin{eqnarray}
\Veff' (v) &=& v \mph = v \Lambda^2 \exp \left[ W_{-1} (\xi)\right] 
= v \Lambda^2 \frac{\xi}{W_{-1} (\xi)}\nn\\
&=& v \frac{ - 2 \frac{(4\pi)^2}{\lambda} \left( m^2 + \frac{1}{2}
      \lambda v^2 \right)}{W_{-1} \left(  \frac{- 2\frac{(4
          \pi)^2}{\lambda} \left( m^2 + \frac{1}{2} \lambda
            v^2\right)}{\Lambda^2} \right) }\label{Vprime}
\end{eqnarray}
We sketch the above for the two cases $m^2 > 0$ and $m^2 < 0$ in
Fig.~\ref{derivative}.
Obviously, the effective potential is minimized at $v=0$ for $m^2 >
0$, and at $v = v_{\mathrm{min}} \equiv \sqrt{\frac{- 2
    m^2}{\lambda}}$ for $m^2 < 0$. The effective potential is valid
only for $v^2 \ll \Lambda^2$, and its behavior for $v \approx \Lambda$
should not be taken at a face value.

Integrating (\ref{Vprime}) with respect to $v$, we obtain the main
result of this paper
\begin{eqnarray}
\Veff (v) &=& \frac{\Lambda^4}{(4 \pi)^2} \frac{1}{8} \xi^2 
\left( \frac{1}{- \frac{1}{2} W_{-1} (\xi)} - \frac{1}{4} \frac{1}{\left( -
      \frac{1}{2} W_{-1} (\xi)\right)^2} \right)
\nn\\
&=& \frac{1}{(4 \pi)^2} \frac{1}{2} \lb \frac{(4
  \pi)^2}{\lambda} \left( m^2 + \frac{1}{2} \lambda v^2 \right)\rb^2
\left( \frac{1}{- \frac{1}{2} W_{-1} (\xi)} - \frac{1}{4}
  \frac{1}{\left( - \frac{1}{2} W_{-1} (\xi)\right)^2} \right)
\end{eqnarray}
where $\xi$ is given by (\ref{xidef}).  Here, we have arbitrarily chosen
the integration constant so that $\Veff$ vanishes at $\xi=0$.  We note
that our result is valid only for $- \xi \ll 1$.

To understand the structure of the effective potential, we introduce a
physical coupling, defined for $m^2 > 0$, by
\begin{equation}
\xph \equiv \frac{1}{- \frac{1}{2} W_{-1} \left( \frac{m^2}{\Lambda^2}
  \ln \frac{\mu^2}{\Lambda^2} \right)}
\label{lambdaph}
\end{equation}
which is an RG invariant.  Especially for $m^2 = \mu^2$, we find
\begin{equation}
\xph = \frac{\lambda}{(4 \pi)^2}
\end{equation}
Looking at the effective potential, we notice that $\frac{1}{-
  \frac{1}{2} W_{-1} (\xi)}$ is $\xph$, where the positive $m^2 +
\frac{\lambda}{2} v^2$ replaces $m^2$, not necessarily positive.
Since the physical coupling admits an asymptotic expansion in powers
of $\lambda > 0$, the effective potential also admits such an
asymptotic expansion.

At the minimum of $\Veff$, we find
\begin{equation}
\frac{d^2 \Veff}{dv^2}\Big|_{\mathrm{minimum}}
 = \lb\begin{array}{l@{\quad}l}
\mph\Big|_{v=0} = m^2 \frac{\lambda_{\mathrm{ph}}}{\lambda}& (m^2 > 0)\\
\mph\Big|_{\xi=0} = 0& (m^2 < 0)\end{array}\right.
\end{equation}
For $m^2 > 0$, this gives the physical squared mass of $\phi^I$ in the
absence of a constant source.  The vanishing for $m^2 < 0$ does not
imply that $\phi^N$ is massless; only $\phi^I\,(I=1,\cdots,N-1)$ are
massless, and $\phi^N$ which is coupled with the auxiliary field
$\alpha$ has a squared mass of order $(- 2 m^2)$.  For $m^2 > 0$, the
fourth order derivative of the effective potential at $v=0$ is not
quite $\lambda_{\mathrm{ph}}$, but given by
\begin{equation}
  \frac{1}{3} \frac{d^4}{dv^4} \Veff (v=0) =
  \frac{\lambda_{\mathrm{ph}}}{1 - \frac{1}{2} \xph} 
\end{equation}
For $m^2 > 0$, the two physical parameters $\xph$ and $\mph$ determine
the effective potential uniquely.

\section{Defending the large $N$ limit}

In \cite{Coleman:1974jh}, two problems were found with the physics of
the large $N$ limit.  One is that the effective potential $\Veff (v)$
is not well defined for $v$ of order $\Lambda$ and larger.  The other
is the presence of a tachyon pole in the propagator of $\phi^N$ in the
broken phase.  Both problems arise due to the neglect of the physical
presence of an UV cutoff.  We resolve them both in the following.

\subsection{$\Veff$ for large $v$}

We have solved (\ref{to-solve}) to obtain the effective action.  Its
right-hand side is monotonically decreasing with $v^2$, but its
left-hand side is monotonically decreasing with $\mph$ only for $\mph
< \e^{-1} \Lambda^2$.  Thus, the one-to-one relationship between $v^2$
and $\mph$ is obtained only for $\mph < \e^{-1} \Lambda^2$.  Because
of this, $\Veff$ is defined only up to $v^2$ of order $\Lambda^2$.  We
find nothing wrong with this, since (\ref{to-solve}) is valid only for
$v^2 \ll \Lambda^2$.

In order to define $\Veff$ for large $v$, we must introduce
non-universality by keeping the neglected terms in the integral
(\ref{integral}).  Using a sharp momentum cutoff, we evaluate
\begin{equation}
\int \frac{d^4 p}{(2 \pi)^4} \frac{1}{p^2 + \mph} \theta (\Lambda_0^2
- p^2) = \frac{1}{(4 \pi)^2} \left[ \Lambda_0^2 + \mph \ln
    \frac{\mph}{\Lambda_0^2} - \mph \ln \left(1 +
        \frac{\mph}{\Lambda_0^2}\right) \right]
\end{equation}
corresponding to $A_4 = \frac{1}{(4 \pi)^2}$ and $B_4 = 0$.  We keep the last
term, which is non-universal.

\begin{figure}[h]
\includegraphics[width=8cm]{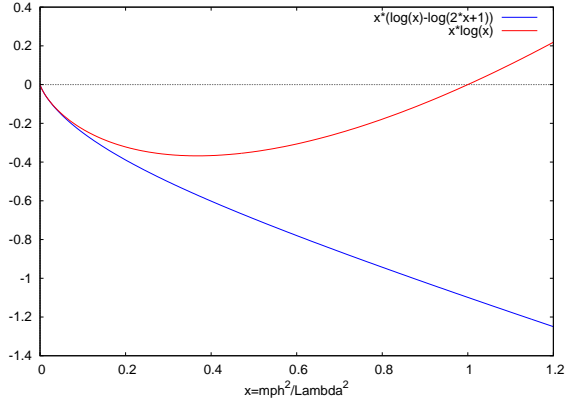}
\caption{We plot the left-hand side of (\ref{one-to-solve}) as a
  function of $x \equiv \frac{\mph}{\Lambda^2}$ for
  $\e^{2 \frac{(4\pi)^2}{\lambda_0}} = 2$.  It is monotonically
  decreasing as opposed to the approximation $x \ln x$ used for
  (\ref{to-solve}).}
\label{large-v}
\end{figure}

The kept term changes (\ref{to-solve}) to
\begin{equation}
\frac{\mph}{\Lambda^2} \left[ \ln \frac{\mph}{\Lambda^2} - \ln \left(1
        + \frac{\mph}{\Lambda^2} \e^{2 \frac{(4 \pi)^2}{\lambda_0}}
    \right)\right] = \xi
\label{one-to-solve}
\end{equation}
where $\xi$ is defined by (\ref{xidef}).  As long as $\lambda_0 > 0$,
the left-hand side is a monotonically decreasing function of $\mph \ge
0$.  (See Fig.~\ref{large-v}.)
For $\mph \gg \Lambda^2$, we obtain
\begin{equation}
- 2 \frac{(4 \pi)^2}{\lambda_0} \mph - \Lambda_0^2
\simeq - 2 \frac{(4 \pi)^2}{\lambda} \left( m^2 + \frac{\lambda}{2}
    v^2 \right)
\end{equation}
where we have used
\begin{equation}
\Lambda_0^2 = \Lambda^2 \e^{- 2 \frac{(4 \pi)^2}{\lambda_0}}
\end{equation}
This gives
\begin{equation}
\Veff (v) \simeq m_0^2 \frac{v^2}{2} + \frac{\lambda_0}{2}
\left(\frac{v^2}{2}\right)^2
\end{equation}
Thus, somewhat as expected, the effective potential for very large $v$
is given by the potential in the bare action.

\subsection{Absence of tachyons in the broken phase}

In the broken phase $m^2 < 0$ of the large $N$ limit, the propagator
of $\phi^N$ is given by
\begin{equation}
\vev{\phi^N (k) \phi^N (-k)} = \frac{\frac{1}{\lambda_0} + \frac{1}{2}
  f\left(\frac{k^2}{\Lambda_0^2}\right)}{k^2 \left(
      \frac{1}{\lambda_0} + \frac{1}{2} 
  f\left(\frac{k^2}{\Lambda_0^2}\right) \right) + \frac{- 2 m^2}{\lambda}}
\label{propagator}
\end{equation}
where we use a sharp cutoff to define
\begin{equation}
f\left(\frac{k^2}{\Lambda_0^2}\right) \equiv \int \frac{d^4 p}{(2
  \pi)^4} \frac{1}{p^2 (p+k)^2} \theta (\Lambda_0^2 - p^2) \theta
(\Lambda_0^2 - (p+k)^2)
\end{equation}
The above propagator is the same as (3.14) of \cite{Coleman:1974jh}.

Note that
\begin{enumerate}
\item $f$ is non-negative, and is decreasing with $\frac{k^2}{\Lambda_0^2}$;
\item $f = 0$ for $k^2 \ge 4 \Lambda_0^2$ due to the sharp UV cutoff.
\end{enumerate}
Since the denominator of the propagator is strictly positive, there is
no tachyonic pole.

We find a tachyonic pole, however, if we use an approximation of $f$.
For $k^2 \ll \Lambda_0^2$, we can approximate $f$ as follows:
\begin{equation}
f \left(\frac{k^2}{\Lambda_0^2}\right) \simeq \frac{1}{(4 \pi)^2}
\left( \ln \frac{\Lambda_0^2}{k^2} + 1 \right)
\end{equation}
If we use this approximate expression, we find a tachyonic pole at
\begin{equation}
k^2 \simeq \e^{2 \frac{(4\pi)^2}{\lambda_0}+1} \Lambda_0^2
\end{equation}
But the approximation of $f$ is invalid for $k^2$ as large as this,
and the pole is indeed absent as we have concluded above by a simple
observation.

\section{Conclusion}

In conclusion, we have obtained the effective potential for the large
$N$ limit of the $O(N)$ linear sigma model explicitly in terms of the
Lambert $W$ function.  

Our use of the Lambert $W$ function is by no means the first
application in quantum field theory.  It has been used to solve
exactly the 2- and 3-loop RG equations in QCD.\cite{Gardi:1998qr,
  Magradze:1998ng, Magradze:1999um} (See also \cite{Nesterenko:2003xb}
for a review of other applications in QCD.)  It has also been applied
for the study of general RG flows in \cite{Curtright:2010hq}, and most
recently by the author to solve generic 2-loop RG equations in
theories with a Gaussian IR fixed point.\cite{Sonoda:2013b}

\appendix*

\section{Lambert $W$ function}

\begin{figure}[h]
\includegraphics[width=8cm]{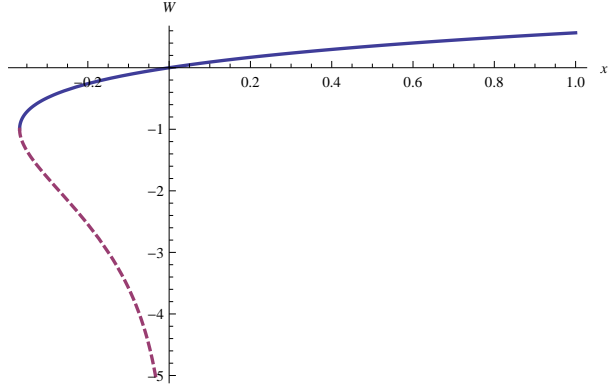}
\caption{The real valued Lambert $W$ function has two branches: the
  upper $W_0$ (solid) and lower $W_{-1}$ (dashed).}
\label{lambert_w}
\end{figure}

The Lambert $W$ function (a.k.a. product logarithm)\cite{wiki:LambertW}
is defined implicitly by
\begin{equation}
W(x) \e^{W(x)} = x
\end{equation}
Restricted to real values, the function has two branches: the upper
$W_0 (x) > -1$ defined for $x \in [- \e^{-1}, +\infty)$ and the lower
$W_{-1} (x) < -1$ for $x \in [- \e^{-1}, 0)$.  (See
Fig.~\ref{lambert_w}.)
For $- x \ll 1$, we obtain the asymptotic expansion
\begin{equation}
W_{-1} (x) = \ln (-x) - \ln \left(- \ln (-x)\right) + \mathrm{O}
\left(\frac{\ln \left(- \ln (-x)\right)}{\ln (-x)}\right)
\end{equation}

Perhaps the best known use of the Lambert $W$ function is for the
distribution function
\begin{equation}
f (x) = \frac{x^n}{\e^x - 1}\quad (x > 0)
\end{equation}
with $n > 0$.  For $x > 0$, it is maximized at $x = n + W_0 (- n \e^{-n})$.

\begin{acknowledgments}
I thank Yannick Meurice for giving me Ref.~\cite{David:1985zz}.
\end{acknowledgments}

\bibliography{lambert}

\end{document}